\documentclass[onecolumn]{article}\makeatletter
\newcommand{\Rmnum}[1]{\expandafter\@slowromancap\romannumeral #1@}
\makeatother
\usepackage{xcolor}
\usepackage{tikz}
\usetikzlibrary{shapes,snakes,backgrounds}
\usepackage{verbatim}
\usepackage{anysize}\marginsize{10mm}{10mm}{10mm}{10mm}

\title{Airfoil Roll Control by Bang-Bang Optimal Control Method with Plasma Actuators}

 \author{
  Qingkai Wei%
    \thanks{
Ph.D. student, Department of Aeronautics and Astronautics; weiqingkai@pku.edu.cn.}\\
  {\normalsize\itshape
   Peking University, Beijing, 100871, People's Republic of China}\\
   and\\
  \ Zhongguo Niu,
  \thanks{
  Engineer, Aerodynamics Research Institute.}
  and
  \ Bao Chen\
  \thanks{
  Senior Engineer, Aerodynamics Research Institute.}\\
  {\normalsize\itshape
   Aviation Industry Corporation of China, Harbin, 150001, People's Republic of China}\\
   and\\
  Xun Huang
  \thanks{
  Associate Professor. Affiliations: (1) Department of Aeronautics and Astronautics; and (2) State Key Laboratory of Turbulence and Complex Systems. AIAA Senior Member. Corresponding author: huangxun@pku.edu.cn. This work has been partially supported by the National Science Foundation Grant of China (90916003) and the Science Foundation of Aeronautics of China (20091571).} \\
  {\normalsize\itshape
   Peking University, Beijing, 100871, People's Republic of China}\\
}

\usepackage{graphics}
\usepackage{graphicx}
\usepackage[]{subfigure}

 \usepackage{cite}
 \usepackage{lettrine}

\begin{document}
\maketitle

\section*{Nomenclature}
\begin{tabbing}
  XXX \= \kill
  $b$  \> \hspace{8mm}\hspace{2mm} = \hspace{4mm} exposed electrodes width, 1mm\\
  $c$  \> \hspace{8mm}\hspace{2mm} = \hspace{4mm} clearance distance of consecutive electrodes, 0mm\\
  $C_z$ \> \hspace{8mm}\hspace{2mm} = \hspace{4mm} aerodynamic force coefficient in $z$-axis, $F_z/(\bar{q}S)$\\
  $C_{l_p}$ \> \hspace{8mm}\hspace{2mm} = \hspace{4mm} roll moment coefficient caused by plasma actuators, \\
  $d$  \> \hspace{8mm}\hspace{2mm} = \hspace{4mm} chord of airfoil, 300mm\\
  $e$  \> \hspace{8mm}\hspace{2mm} = \hspace{4mm} insulated electrodes width, 4mm\\
  $f $  \> \hspace{8mm}\hspace{2mm} = \hspace{4mm} clearance distance of consecutive plasma actuators, 6mm\\
  $F_z$ \> \hspace{8mm}\hspace{2mm} = \hspace{4mm} aerodynamic force in $z$-axis, N\\
  $(I_x,I_y,I_z)$  \> \hspace{8mm}\hspace{2mm} = \hspace{4mm}  the moment of inertia, $kg\cdot \rm m^2$\\
  $(I_{xy},I_{xz},I_{yz})$  \> \hspace{8mm}\hspace{2mm} = \hspace{4mm} product of inertia, $kg\cdot \rm m^2$\\
  $L$  \> \hspace{8mm}\hspace{2mm} = \hspace{4mm} total roll moment, $N\cdot m$\\
  $L_0$  \> \hspace{8mm}\hspace{2mm} = \hspace{4mm} aerodynamic roll moment, $N\cdot m$\\
  $L_p$  \> \hspace{8mm}\hspace{2mm} = \hspace{4mm} plasma-induced roll moment, $N\cdot m$\\
  $O$  \> \hspace{8mm}\hspace{2mm} = \hspace{4mm} coordinate origin\\
  $(p,q,r)$  \> \hspace{8mm}\hspace{2mm} = \hspace{4mm} angular velocity in body axes, deg/s\\
  $\bar{q}$ \> \hspace{8mm}\hspace{2mm} = \hspace{4mm} dynamic pressure, $\frac{1}{2}\rho U_{\infty}^2$\\
  $Re$ \> \hspace{8mm}\hspace{2mm} = \hspace{4mm} Reynolds number based on $d$ and $U_{\infty}$\\
  $S$ \> \hspace{8mm}\hspace{2mm} = \hspace{4mm} area of the measuring section, $\rm mm^2$\\
  $U_{\infty}$ \> \hspace{8mm}\hspace{2mm} = \hspace{4mm} speed of freestream, m/s\\
  $Ws$ \> \hspace{8mm}\hspace{2mm} = \hspace{4mm} wind span of airfoil, 3000mm\\
$\alpha$  \> \hspace{8mm}\hspace{2mm} = \hspace{4mm} angle of attack, deg\\
  $\rho$ \> \hspace{8mm}\hspace{2mm} = \hspace{4mm} air density, 1.225$\rm kg/m^3$\\
  $(\phi,\theta,\psi)$  \> \hspace{8mm}\hspace{2mm} = \hspace{4mm} Euler angles, deg\\
 \textit{Subscripts}\\
  $A_{T}$    \> \hspace{8mm}\hspace{2mm} = \hspace{4mm} target value of A\\
 \end{tabbing}

\section{Introduction}
\lettrine[nindent=0pt]{P}{LASMA} actuators, operating in atmospheric pressure air, have attracted increasing research interest in aerospace over the past two decades\cite{roth2000electrohydrodynamic,moreau2007airflow}. Most previous publications have focused on flow control applications that demonstrate the capability of plasma actuators and uncover, at least partially, the related fluid mechanics. In the present paper, we go one step further, studying the integration of flight control and flow control using plasma actuators, which constitutes the main contribution of this paper. As a demonstration, rolling maneuver of an airfoil was controlled only using plasma actuators, which could save mechanical moving parts of ailerons. In addition, an optimal flight controller was designed and demonstrated in simulations, taking account of flow control characteristics of plasma actuators. The proposed control method can also be considered for other flow control applications with plasma actuators.

Various plasma actuators have been developed to address various flow control issues, which include high speed flow control using localized arc filament plasma actuators\cite{kim2009active} and surface impulse discharges\cite{gnemmi2008feasibility}, flat plate boundary layer flow control by non-thermal direct current (DC) corona\cite{moreau2006effect} and aerodynamic-generated noise control with glow discharges\cite{huang2010plasma}. A comprehensive review of those plasma actuators can be found in the literature\cite{moreau2007airflow} and references therein. In this work, we adopted dielectric barrier discharge (DBD) plasma actuators\cite{enloe2004mechanisms} to control low speed aerodynamics and, going one step further, to control flight maneuver. A similar investigation has been conducted in a recent work\cite{seifert2010roll}, but using piezo-fluidic actuators. Satisfactory flight control performance has been shown in interesting flight tests\cite{seifert2010roll}. The present work differs from previous publications, focusing on the development of optimal flight control methodology for DBD plasma actuators.

\begin{figure}[htbp]
     \centering
     \subfigure[]
       {\label{DBDskeme}
     \includegraphics[height=40mm]{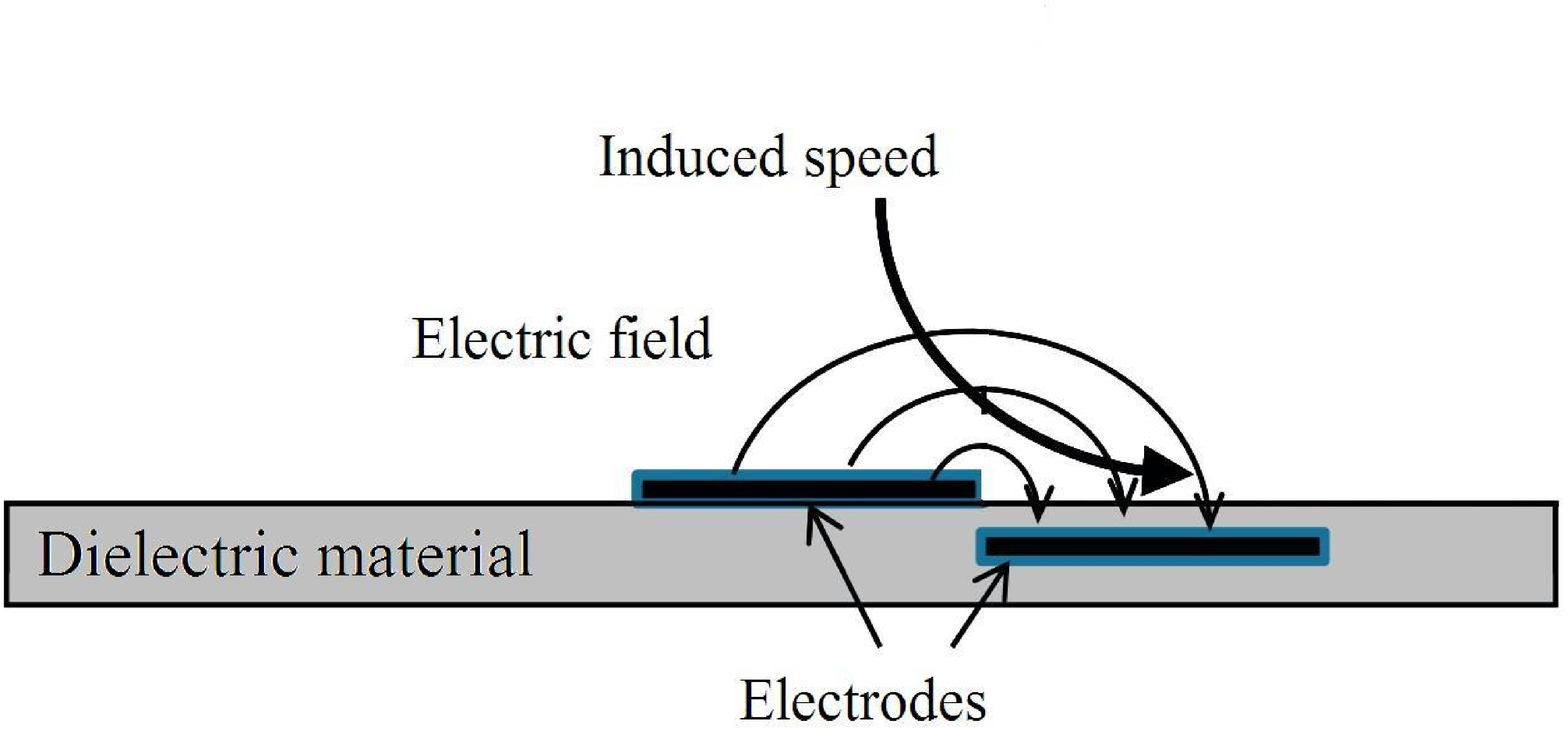}}
     \subfigure[]
       {\label{ACVoltage}
      \includegraphics[height=60mm]{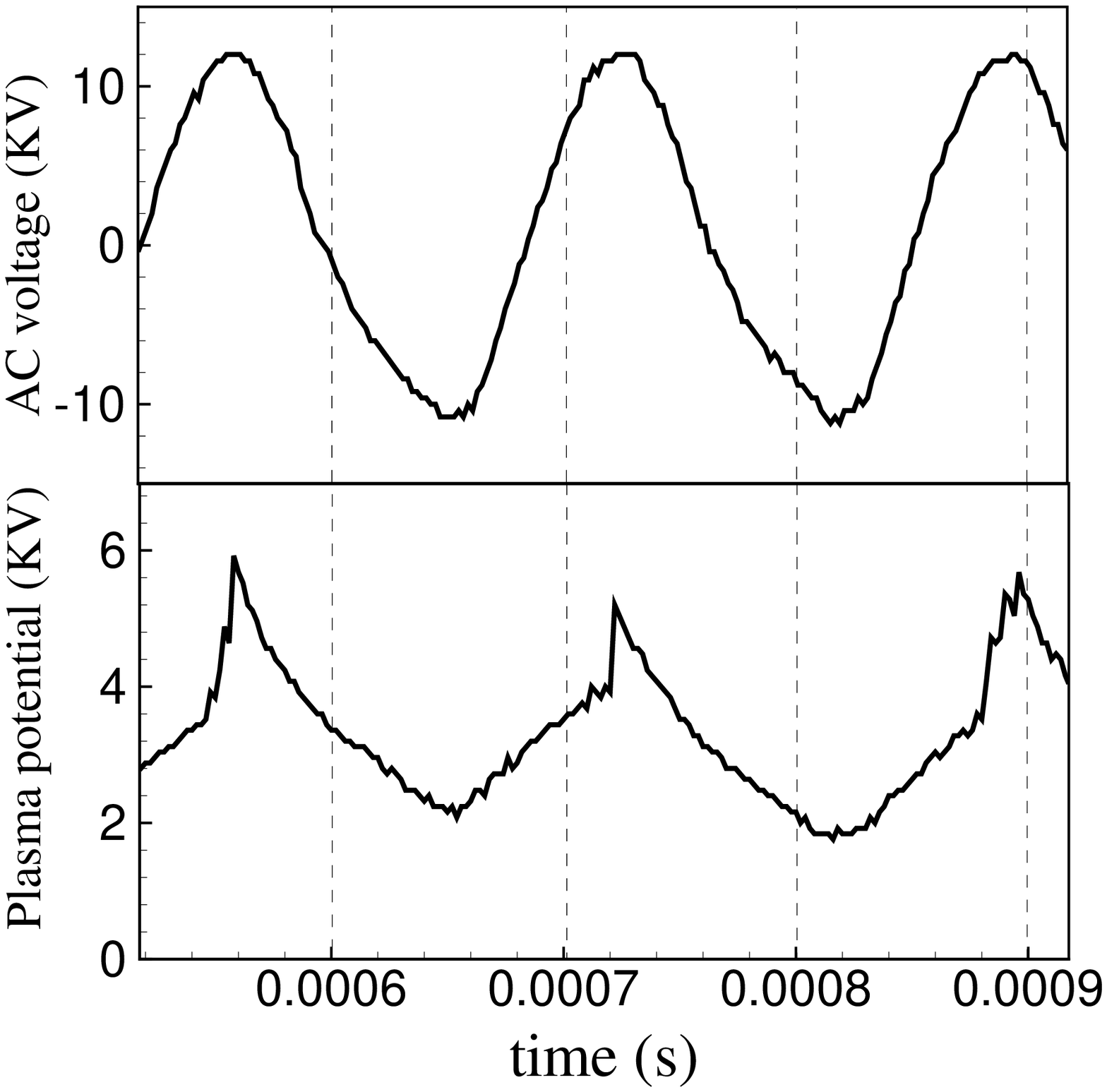}}
     \caption{The plasma actuator, where (a) the schematic of a DBD plasma actuator;\cite{huang2010broadband} and (b) the AC voltage applied to the electrodes (top) and the plasma potential (bottom) measured in the plasma at 10 mm from the exposed electrode.}\label{DBDs}
   \end{figure}

The principle of using DBD plasma actuators for flow control is not new. Figure \ref{DBDskeme} shows that a DBD plasma actuator normally consists of two electrodes isolated by a dielectric material. Potential candidates of the dielectric material include silicon rubber and flame retardant. An alternating current (AC) power supply (Fig. \ref{ACVoltage}) is applied to the two electrodes, between which glow discharges are generated, leading to the momentum transfer from charged nitrogen/oxygen particles to the local neutral gas through collisions. The collective effect is an induced fluid motion along the surface of the dielectric material, from the exposed electrode to the insulated electrode. The induced fluid motion develops vertical structures and manipulates local and global fluid mechanisms\cite{huang2008streamwise}. Detailed discussion of the related plasma and aerodynamic physics can be found in the literature\cite{roth2000electrohydrodynamic}. 

In previous works, deployable flow effectors have been adopted on the upper surface of an NACA 0020 airfoil to delay flow separation\cite{patel2003active}. It has been shown that the stall angle can be increased from 18deg to 20deg. In addition to aerodynamic control, one natural extension is to control airfoil flight dynamics using plasma actuators. In particular, the lift modification due to plasma can generate desired pitching and rolling moments. As a result, it is possible to save the mechanical moving parts for aileron and flaps. However, feedback control system has to be considered along with the plasma active flow control to achieve optimal performance. A variable structure feedback control system has been proposed previously for cavity flow-induced noise control using glow discharges\cite{huang2008variable}. The fundamental idea behind that feedback control case was to adjust the authority of plasma actuation to a required strength, according to the feedback measurements and the variable structure model. The modulation of plasma actuation in real-time\cite{huang2007atmospheric}, however, needs complex electrical circuits and generates serious electromagnetic pollutions. A simpler configuration of plasma actuators that only has two working states (on and off) was considered in this work. A bang-bang controller, also known as an on-off controller that abruptly changes actuation between upper and lower bounds, was used to feedback control the roll of NACA 0015 airfoil. The bang-bang control method is expected to generate less electromagnetic interference and is, inherently optimal in terms of its capability to achieve control objective in minimum time.

It can be seen from Fig. \ref{ACVoltage} that the AC input and the plasma potential have oscillations, which suggest the \lq\lq on\rq\rq\ state of a plasma actuator changes over time. It has been discovered in our previous work\cite{huang2010broadband} that the maximum amplitude of the plasma-induced speed rapidly varies between 6m/s and 8m/s, leading to an oscillating lift force and force moment. The situation will be worsened for practical cases with time varied power supply at various atmospheric environments. This practical issue was considered and the disturbance rejection capability of the control method was studied in this work. Moreover, most DBD plasma actuators have a quite limited control performance for high speed flow cases. Different plasma actuators, such as impulse discharges\cite{gnemmi2008feasibility} and corona\cite{moreau2006effect} have been suggested for such cases. The attention of this paper focuses on low speed aerodynamic cases (around 15m/s) and can find applications in unmanned aerial vehicles. However, it is worthwhile to emphasize that the proposed control method is generic and could be applicable for applications with other plasma actuators.

The following of this paper is organized as follows. Section \Rmnum{2} describes experimental apparatus that achieve aerodynamic data for the next flight control simulation. The bang-bang control method was briefly introduced in Sec. \Rmnum{3} to complete the paper. A bang-bang controller was thereafter designed particularly for the airfoil roll control case with DBD plasma actuators. The integration of flight control and active flow control was simulated in Sec. \Rmnum{4}, where the oscillations of plasma actuations and the uncertainties in aerodynamic data were considered and addressed. A brief summary was provided in the end of this paper.

\section{Experimental Apparatus}
\begin{figure}[htbp]
     \centering
     \includegraphics[width=140mm]{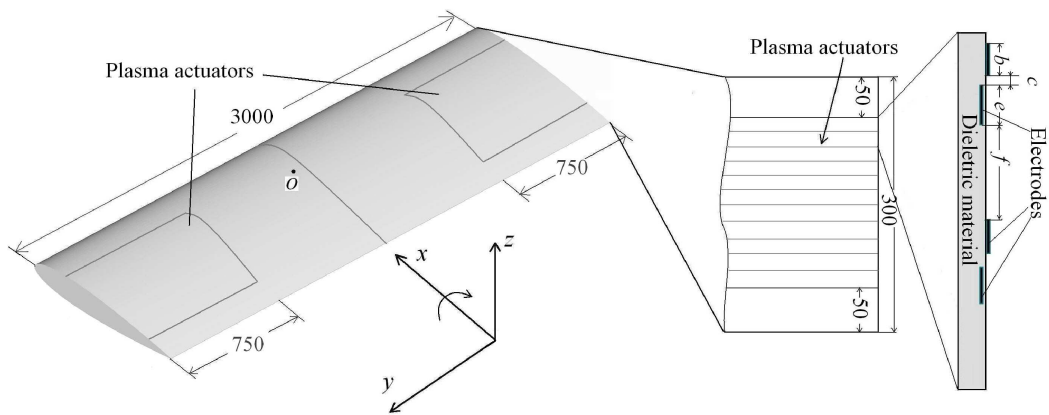}
     \caption{The schematic of the NACA 0015 airfoil and the layout of plasma actuators for the roll control simulation case (dimensions are in millimeters).}\label{model}
  \end{figure}

Roll control of an NACA 0015 airfoil (0.3m chord, 3m span) with plasma actuators was studied in this work. Figures \ref{model} shows the body axes used throughout the rest of the paper. The origin $O$ of the coordinates is located at the center of gravity of the airfoil. Plasma actuators are placed on the upper surface. The span of the plasma actuators is 0.75m, evenly covering the left and right wings (Fig. \ref{model}). Figure \ref{model} shows an enlarged view of one plasma area, which is composed of nineteen consecutively spaced plasma actuators. The structure and configuration of each plasma actuator are similar to those in Fig. \ref{DBDskeme}. All nineteen plasma actuators are parallel circuits and can be simultaneously switched on/off to a high voltage AC power supply. In roll control case, the plasma actuators on either left or right wing can be independently switched on/off. 

  \begin{figure}[htbp]
     \centering
     \includegraphics[width=90mm]{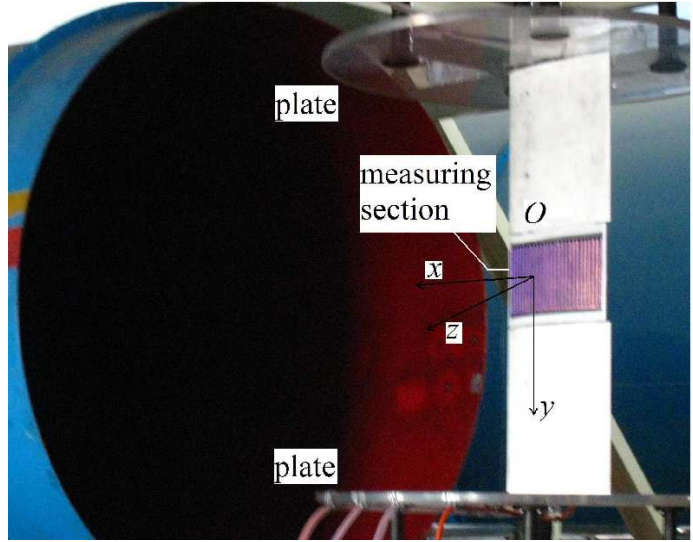}
     \caption{The setup in the wind tunnel with plasma glow discharges.}\label{apparatus}
  \end{figure}

The aerodynamic data requested by flight control was achieved by conducting experiments in the FL-5 wind tunnel at Aerodynamics Research Institute. The FL-5 low-speed wind tunnel has an open-return circuit design. The open testing section is round with a diameter of 1.5m. The wind speed ranges from 0m/s to 53m/s. Figure \ref{apparatus} shows the experimental setup.

An NACA 0015 airfoil (0.3m chord, 0.9m span) was manufactured by Teflon and installed in the open test section. The rotating plates that can rotate within $\pm24$deg are used to hold the model and test equipments as well as to maintain a good flow quality. Only the central section (0.2m span) of the model is covered with plasma actuators. As a result, the potential influence on the plasma flow control from the three-dimensional fluid and boundary flow local to the rotating plates can be omitted. The thickness of the dielectric material (epoxy polymer) is 1.5mm. The surface of the NACA 0015 airfoil can be etched to smoothly contain the plasma actuators. The electrodes are made by copper. The width of the exposed copper electrodes is $b=1$mm. The width of the insulated electrodes is $e= 4$mm. The clearance distance $c$ between two electrodes is zero. The clearance distance between two plasma actuators is $f=6$mm. It should be pointed out that the above values are empirically chosen, reflecting a tradeoff between the experimental geometry limitation and plasma performance. It can be seen that the simulation setup shares the same geometrical setup for plasma actuators. Hence, the plasma induced roll momentum can be derived from the aerodynamic force measurements, although the experimental model is different from the simulation model for the convenience of the experiments. The force and moment coefficients of the simulation model, with and without plasma actuation, were respectively calculated based on experimental results, taking geometry difference into account. The AC voltage applied is 22kV, 180W at 4.7kHz. The following simulations adopt experimental results achieved at 15m/s. The corresponding Reynolds number with respect to the chord length is about $10^6$.

    \begin{figure}[htbp]
     \centering
     \subfigure[]
       {\label{Cz}
     \includegraphics[width=75mm]{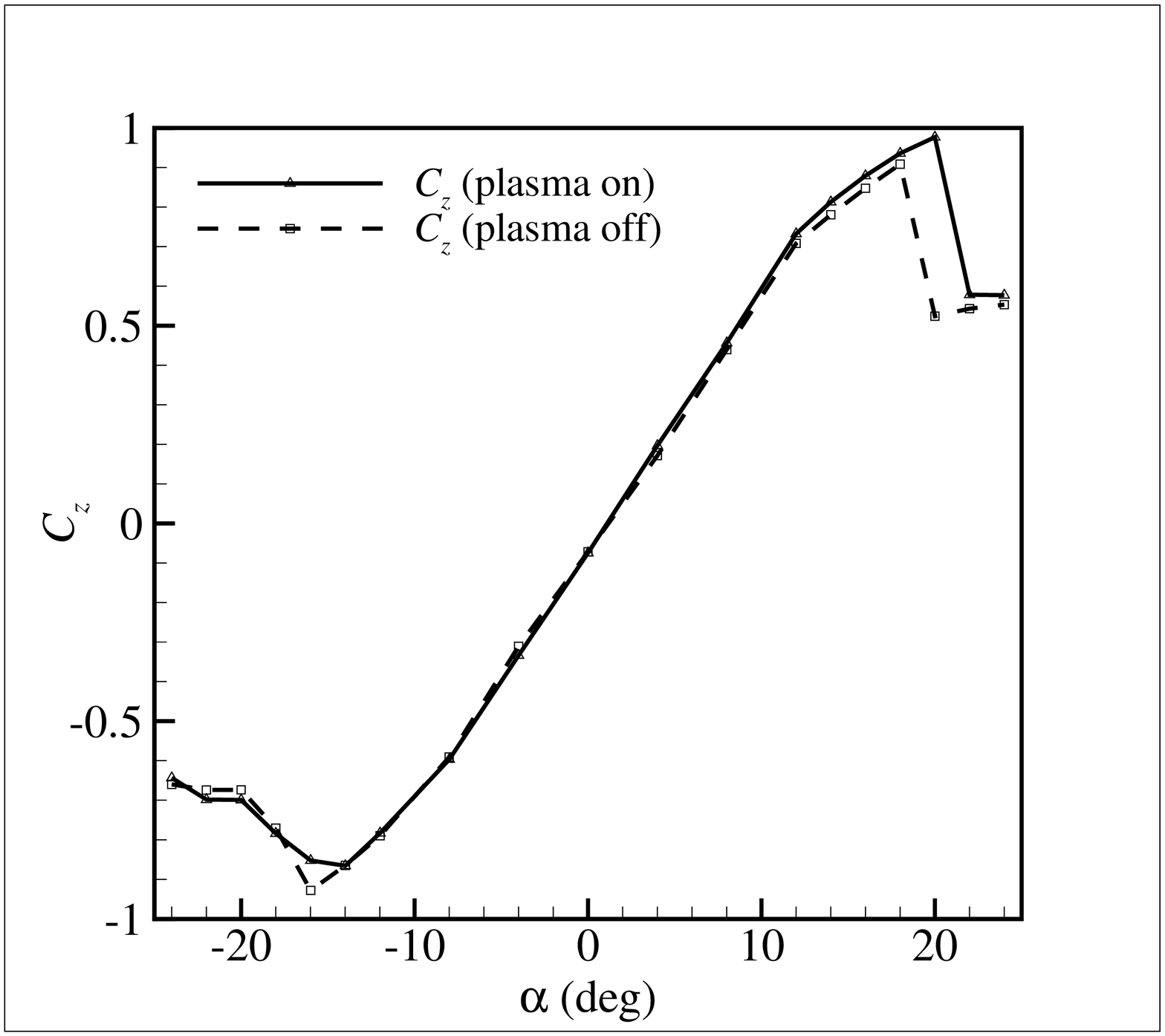}}
     \subfigure[]
       {\label{Clp}
      \includegraphics[width=75mm]{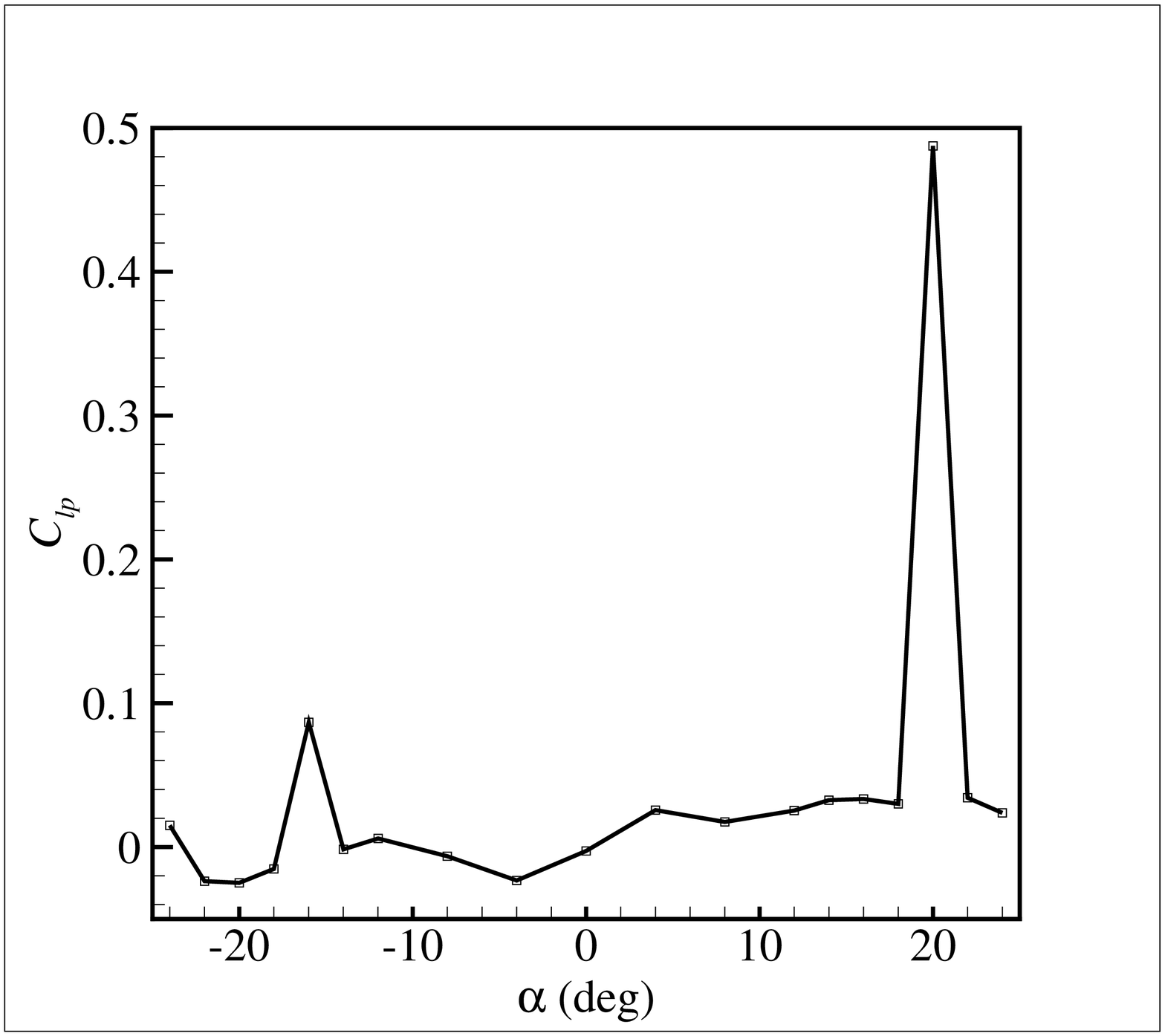}}
     \caption{(a) Experimental results of $C_z$ at $U_{\infty}=$15m/s, where the corresponding Reynolds number is $10^6$; and (b) the related plasma-induced roll moment coefficient in the simulations.}\label{ExperimentData}
   \end{figure}

The experimental measurements of aerodynamic force are shown in Fig.  \ref{Cz}. The force coefficient $C_z=F_z/(\bar{q}S)$, where $F_z$ is the force along the $z$-axis, $S$ is the area of the airfoil, $\bar{q}$ is dynamic pressure that is $\rho U_{\infty}^2/2$, $\rho$ is the air density, $U_{\infty}$ is the freestream speed. It can be seen that the amplitude of  $C_z$ increases almost linearly along with the increase of $\alpha$. The $C_z$ quickly drops beyond the so-called stall angle. Plasma actuation can slightly increase stall angle by 3deg. On the other hand, the increase of $C_z$ due to plasma actuation at low angles of attack is quite small. For example, we can only have 3.56\% increase (from 0.705 to 0.73) in $\left|C_z\right|$ at $\alpha=12$deg. Figure \ref{Clp} shows the plasma-induced roll moment coefficient that is used in the simulations, where plasma actuators in the left airfoil are activated and the right ones are deactivated. The relatively large roll moment around 20deg is caused by the increased stall angle using plasma actuation. The roll moments at other angles are small. However, it can be seen in the following simulations that even such small roll moments can achieve good roll control performance. 

\section{Bang-Bang Controller Design}
In control theory, bang-bang control is well known as the minimum-time optimal feedback control method, whose control inputs are constrained to only two levels\cite{lewis1995optimal}. In particular, the roll dynamics of the airfoil can be described for the airfoil roll control case by
\begin{eqnarray}
 \nonumber L&=&I_x\dot{p}-(I_y-I_z)qr+I_{xy}(pr-\dot{q})-I_{xz}(pq+\dot{r})+I_{yz}(r^2-q^2),\\
  L&=&L_0+L_p,\\
 \nonumber p&=&\dot{\phi}-(\mathrm{sin}\theta)\dot{\psi},
\end{eqnarray}
 \noindent where $(I_x,I_y,I_z)$ is the moment of inertia in body axes, $I_{xy}$ is the product of inertia about $ox$ and $oy$ axes, $I_{xz}$ is the product of inertia about $ox$ and $oz$ axes, $I_{yz}$ is the product of inertia about $oy$ and $oz$ axes. The angular velocities are represented by $(p,q,r)$. The Euler angles of the airfoil about the flat earth are $(\phi,\theta,\psi)$. $L$ is the total roll moment, which consists of aerodynamic roll moment $L_0$ and plasma-induced roll moment $L_p$.

The produces of inertia $I_{xy}=I_{yz}=0$, as the airfoil is symmetric about the $oxz$ plane and the airfoil has a uniform mass distribution. In addition, the pitch and yaw motions are excluded for simplicity, i.e. $q\equiv0$, $r\equiv0$, $\psi\equiv0$, $\theta\equiv0$. As a result, $\dot{\phi}=p$, where $p$ is the roll angular velocity in body axes.

It is implicitly assumed that the airfoil is already trimmed without plasma actuation. The plasma-induced moment is $L_p=M$ in case the plasma actuators on the left wing are activated. Similarly, $L_p=-M$ if the plasma actuators on the right wing are activated. The scenario that all plasma actuators on both wings are activated was not considered in this work. In summary, the roll dynamics can be simplified to
\begin{equation}
  \ddot{\phi}=\pm M/I_{x}.
\end{equation} 

A state space model can be accordingly formulated as

\begin{equation}\label{mathematicmodel}
  \left[\begin{array}{ll}
  \dot{\phi}\\
  \ddot{\phi}
  \end{array}\right]
  =
  \left[
  \begin{array}{ll}
  0&1\\0&0
  \end{array}
  \right]
  \left[
  \begin{array}{ll}
  \phi\\ \dot{\phi}
  \end{array}
  \right]
  +
  \left[
  \begin{array}{ll}
  0\\ M/I_x
  \end{array}
  \right]
  u,\
  u=\pm1,
\end{equation}
\noindent where the control input $u$ is the sign of plasma-induced roll moment.

\begin{figure}[htbp]
     \centering
     \subfigure[]
       {\label{bangbang1}
\begin{tikzpicture}[domain=0:4,scale=0.4]
    \fill[black] (0,0) circle (2pt) node [below right] {\large$O$};
    \draw[line width=1pt][->] (-9,0) -- (9,0) node[right] {\large $\phi$};
    \draw[line width=1pt][->] (0,-5) -- (0,5) node[above] {\large $\dot{\phi}$};
    
    \draw[color=black][line width=1pt] plot ({\x*\x/3},-\x);
    \draw[line width=1pt][->] plot ({\x*\x/3}, \x);
    \draw[line width=1pt] plot ({\x*\x/3+4},-\x); 
    \draw[line width=1pt][->]  plot ({\x*\x/3+4}, \x);
    \draw[line width=1pt] plot ({\x*\x/3+2},-\x); 
    \draw[line width=1pt][->]  plot ({\x*\x/3+2}, \x);
    \draw[line width=1pt] plot ({\x*\x/3-4},-\x);  
    \draw[line width=1pt][->]  plot ({\x*\x/3-4}, \x);
    \draw[line width=1pt] plot ({\x*\x/3-2},-\x);  
    \draw[line width=1pt][->]  plot ({\x*\x/3-2}, \x);
    \draw (8,1.5) node[below left] {\large $u=+1$};
    
    \draw[line width=1pt][->] plot ({-\x*\x/3},-\x);
    \draw[line width=1pt] plot ({-\x*\x/3}, \x);
    \draw[line width=1pt][->]  plot ({-\x*\x/3+4},-\x);
    \draw[line width=1pt] plot ({-\x*\x/3+4}, \x);
    \draw[line width=1pt][->]  plot ({-\x*\x/3+2},-\x);
    \draw[line width=1pt] plot ({-\x*\x/3+2}, \x);
    \draw[line width=1pt][->]  plot ({-\x*\x/3-4},-\x);
    \draw[line width=1pt] plot ({-\x*\x/3-4}, \x);
    \draw[line width=1pt][->]  plot ({-\x*\x/3-2},-\x);
    \draw[line width=1pt] plot ({-\x*\x/3-2}, \x);
    \draw (-5,1.5) node[below left] {\large $u=-1$};
\end{tikzpicture}}
     \hspace{1cm}
     \subfigure[]
       {\label{bangbang2}
\begin{tikzpicture}[domain=0:4.5,scale=0.4]
    \draw[line width=1pt][->] (-1,0) -- (11,0) node[right] {\large$\phi$};
    \draw[line width=1pt][->] (0,-5) -- (0,5) node[above] {\large$\dot{\phi}$};
    \draw[line width=1.5pt,very thick,dashed,gray]  plot ({\x*\x/3+6},-\x);
    \draw[line width=1.5pt,very thick,dashed,gray]  plot ({-\x*\x/3+6},\x);

    \fill[black] (0,0) circle (2pt) node [below right] {\large$O$};
    \fill[black] (3,3) circle (2pt) node [above] {\large$A$};
    \fill[black] (6.1,0) circle (2pt) node [below right] {\large$B(\phi_T,0)$};
    \draw (5,2.7) node[below right] {\large on-off line};
            \path (0,0) node (o){}
                (3,3) node (a){}
                (6,0) node (b){};
    \draw[line width=1.5pt,->] (o).. controls (0,1.1) and (2,2.7)..(a);
    \draw[line width=1.5pt,->] (a).. controls (4,2.7) and (6,1.1)..(b);    
\end{tikzpicture}}
     \caption{Schematics of bang-bang control, where (a) phase line with $u=\pm1$; (b) the on-off line for roll control.}\label{bangbang}
   \end{figure}
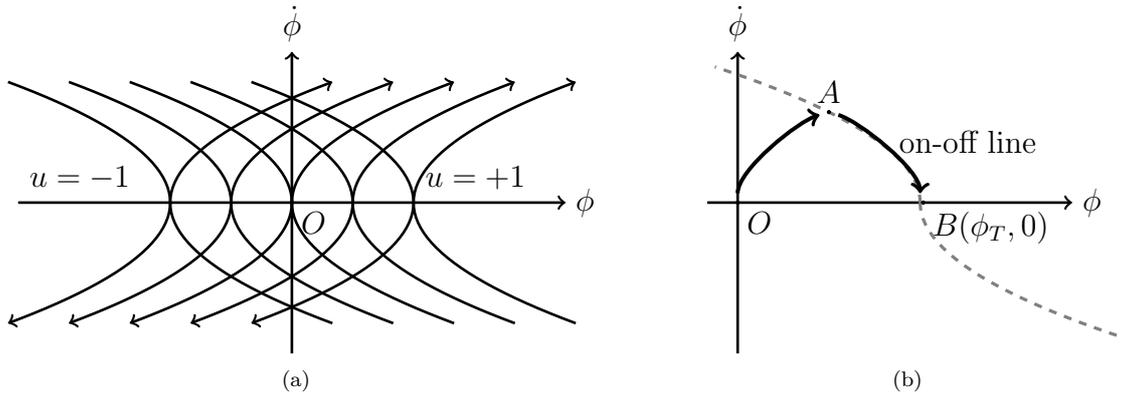

With no loss of generality, the initial state of the airfoil is assumed at $[\phi(0), \dot{\phi}(0)]=[0, 0]$. The objective of the roll control is  $[\phi(t),\dot{\phi}(t)]=[\phi_T, 0]$, where $\phi_T$ is the target roll angle. It is straightforward to achieve
\begin{equation}\label{state}
\left\{
  \begin{array}{l}
     \dot{\phi}=\displaystyle \frac{M}{I_x} t,\\
     \phi=\displaystyle \frac{M}{2I_x}t^2+ C_1,
  \end{array}
  \right. \rm for\ u=1;
  \left\{
  \begin{array}{l}
     \dot{\phi}=-\displaystyle \frac{M}{I_x}t,\\
     \phi=-\displaystyle \frac{M}{2I_x} t^2+ C_1,
  \end{array}
  \right. \rm for\ u=-1,
\end{equation}
\noindent where $C_1$ is a constant value. Figure \ref{bangbang1} shows the phase line $\phi=\pm \frac{1}{2M/I_x}(\dot{\phi})^2+C_1(u=\pm 1)$, which is achieved by eliminating the variable $t$ in Eqs. (\ref{state}).

As shown in Fig. \ref{bangbang2}, the initial state $[\phi(0), \dot{\phi}(0)]=[0, 0]$ is at point $O$. The control objective $[\phi(t),\dot{\phi}(t)]=[\phi_T, 0]$ at $B$. As a result, the initial control input should be set to $u=+1$ to turn the initial state $O$ along the phase line $\phi= \frac{1}{2M/I_x}\dot{\phi}^2$ to state $A$, which is the intersection on the other phase line $\phi=-\frac{1}{2M/I_x}\dot{\phi}^2+\phi_T\ (\dot{\phi}>0)$. The control input switches to $u=-1$ once $A$ is reached. The state thereafter advances to the target state $B$$[\phi_T, 0]$.

A so-called on-off line (the dashed line in Fig. \ref{bangbang2}) only depends on $\phi_T$ and can be formulated as 
\begin{equation}
\phi=-\frac{1}{2M/I_x}\dot{\phi}\vert \dot{\phi}\vert +\phi_T.
\end{equation}

In summary, for any initial solution, the desired control input $u$ is:
\begin{equation}\label{criterion}
  u=
  \left\{
  \begin{array}{l}
  +1;(\phi+\displaystyle  \frac{1}{2M/I_x}\dot{\phi} \vert \dot{\phi} \vert-\phi_T<0)\ \mathrm{or}\ (\phi+ \displaystyle \frac{1}{2M/I_x}\dot{\phi} \vert \dot{\phi} \vert-\phi_T=0,\ \dot{\phi} <0)\\
   -1;(\phi+\displaystyle \frac{1}{2M/I_x}\dot{\phi} \vert \dot{\phi} \vert-\phi_T>0)\ \mathrm{or}\ (\phi+ \frac{1}{2M/I_x}\dot{\phi} \vert \dot{\phi} \vert-\phi_T=0,\ \dot{\phi} >0)
    \end{array}
    \right..
\end{equation}

\section{Simulation and Discussion}
\begin{figure}
     \centering
     \includegraphics[width=140mm]{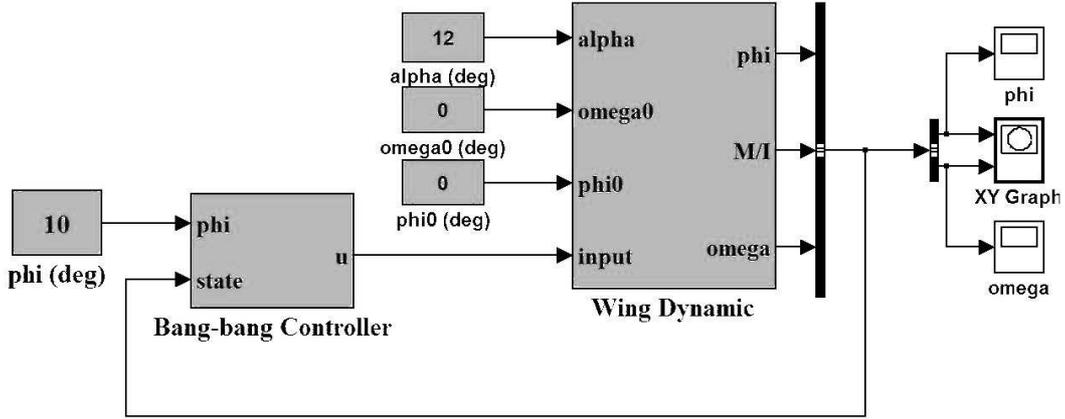} 
      \caption{The simulation case in MATLAB.}\label{Smodel}
    \end{figure}

Numerical simulations were conducted to demonstrate airfoil roll control using plasma actuators. Figure \ref{Smodel} shows the implementation of the simulation case in MATLAB. The block of bang-bang controller implements Eqs. (\ref{criterion}). The wing dynamic block implements the airfoil roll dynamics. The moment of inertia $I_x$ is $4.05\mathrm{kgm^2}$, given the span length (3m) and the uniformly distributed mass ($5.4$kg) with the density of $200\mathrm{kg/m^3}$. The dynamic pressure is $137.8$Pa, given the air density $1.225\mathrm{kg/m^3}$ and mean flow velocity $15\mathrm{m/s}$.
    
     \begin{figure}
     \centering
     \subfigure[]
       {\label{resulta1}
       \includegraphics[height=50mm]{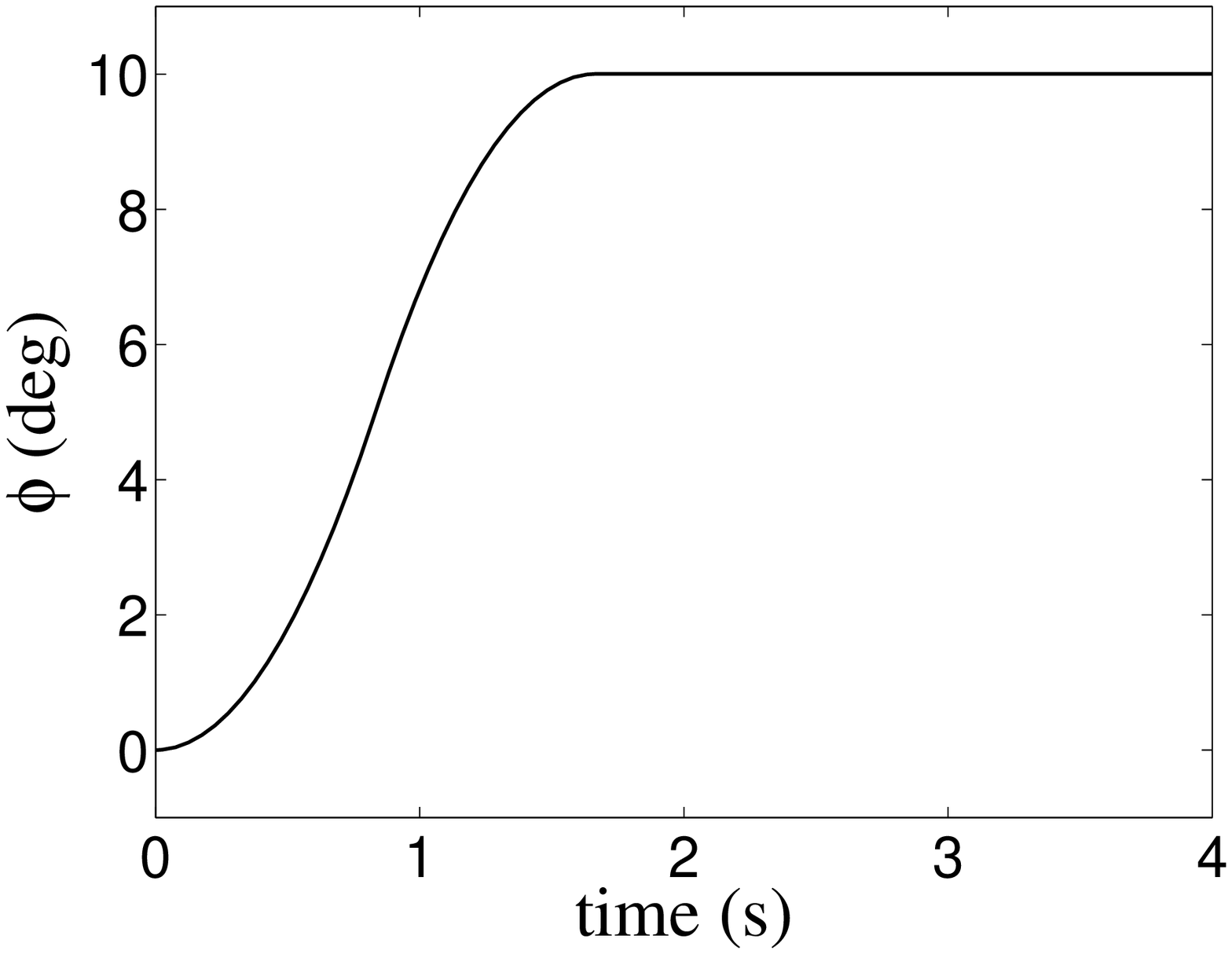}}
     \subfigure[]
       {\label{resulta2}
       \includegraphics[height=50mm]{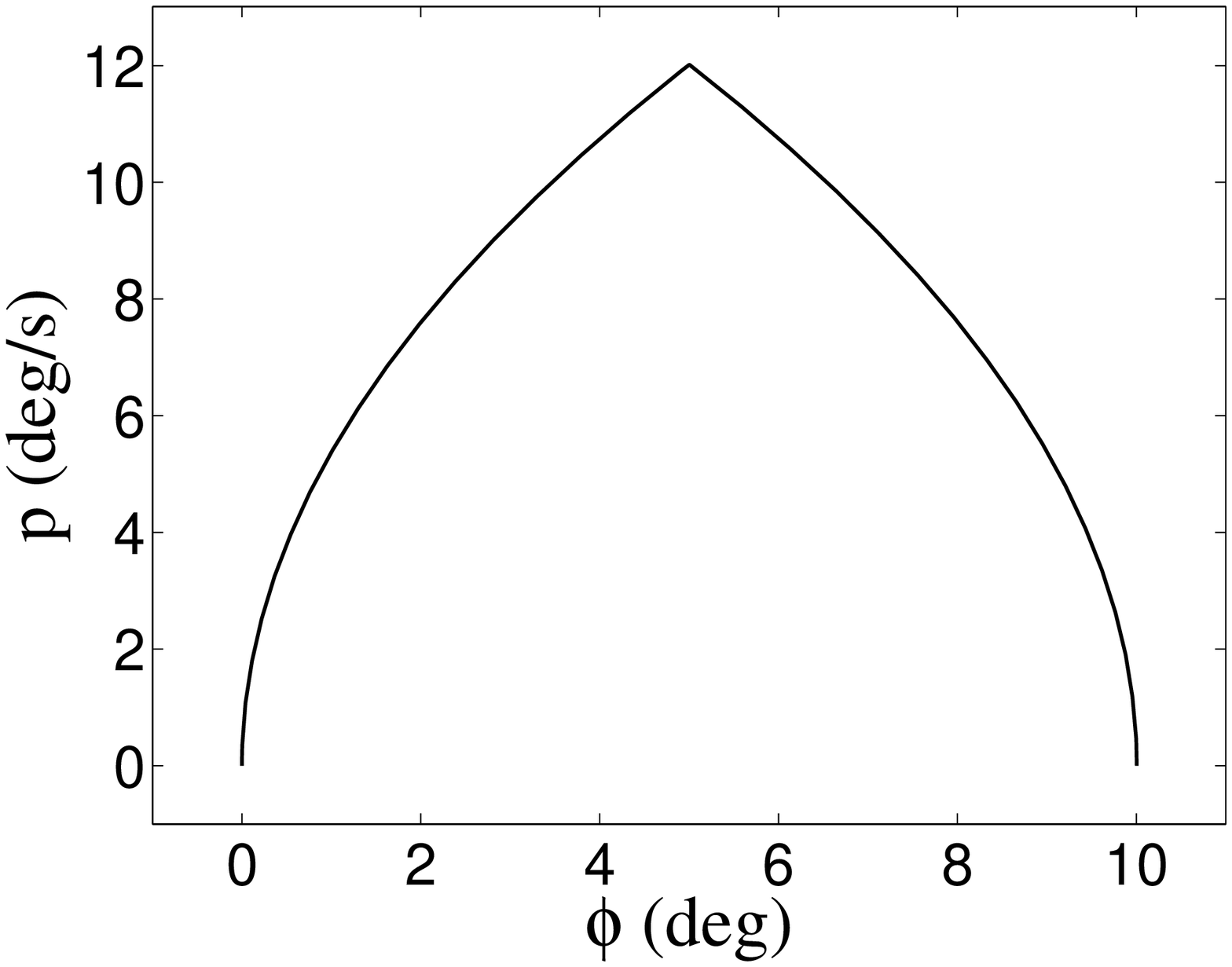}}\\
     \caption{Dynamics during the airfoil rolls from 0deg to 10deg at $\alpha=12$deg and $U_{\infty}=15$m/s, where (a) roll angle dynamics and (b) phase line.}\label{nominal}
     \end{figure}
\begin{figure}
     \centering
     \includegraphics[width=75mm]{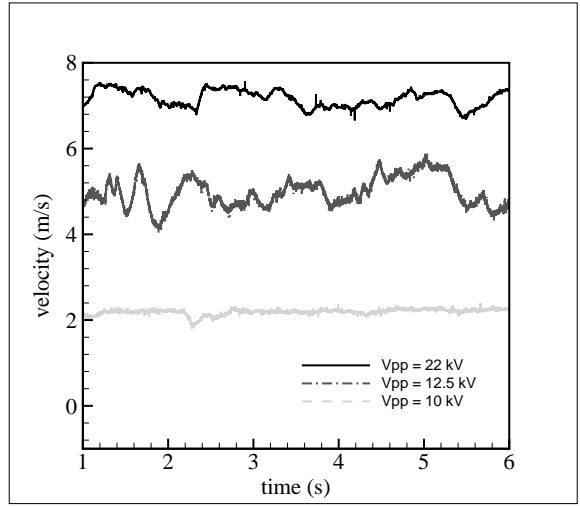} 
      \caption{Measurements of plasma-induced velocities in the stationary atmospheric pressure air.}\label{Plasma_velocity}
    \end{figure}

Figure \ref{nominal} shows the simulation results for $\phi_T=10$deg, where the angle of attack $\alpha=12$deg, at which it is implicitly assumed that the airfoil is already trimmed. It can be seen that plasma-induced moments roll the airfoil to the target angle in about 1.5s. Almost no overshoot can be found. The satisfactory control results with the bang-bang control method may be arguable as the plasma-induced aerodynamic inputs are varying over time (as suggested by Fig. \ref{ACVoltage}).  Figure \ref{Plasma_velocity} is the plasma-induced velocity measured at 20mm downstream from the exposed electrodes\cite{huang2008streamwise}. It can be seen that the plasma-induced velocities oscillate at high voltage input (Vpp=22kV and 12.5kV). The variance is approximately 20\%.

\begin{figure}
     \centering
     \subfigure[]
       {\label{MC11}
       \includegraphics[height=50mm]{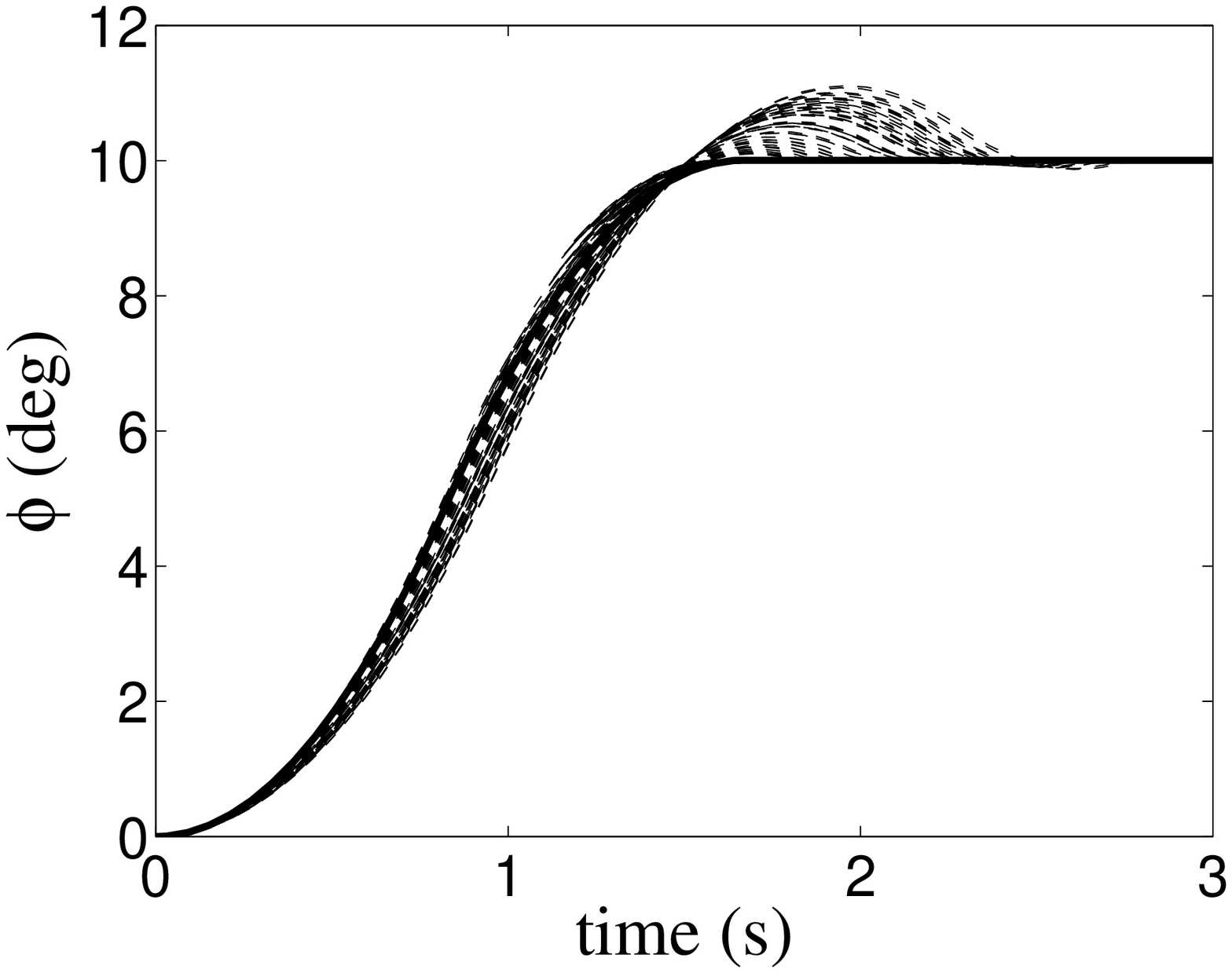}}
     \subfigure[]
       {\label{MC12}
       \includegraphics[height=49mm]{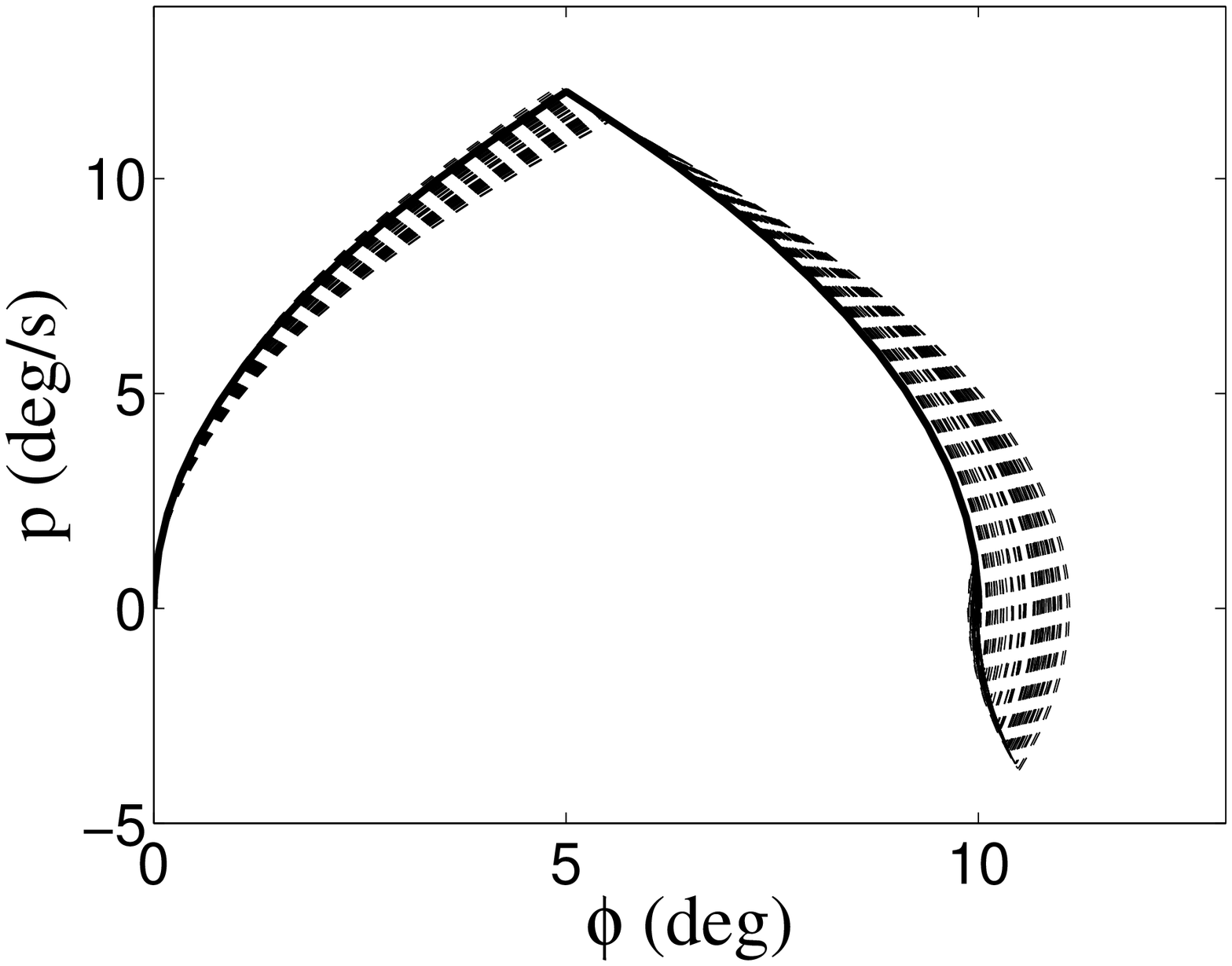}}\\
     \caption{Monte Carlo simulation results of  roll control from 0deg to 10deg, where the plasma-induced roll momentum is randomly chosen, where (a) roll angle dynamics and (b) phase line.}\label{MC1}
     \end{figure}

\begin{figure}
     \centering
     \subfigure[]
       {\label{MC2a}
       \includegraphics[height=55mm]{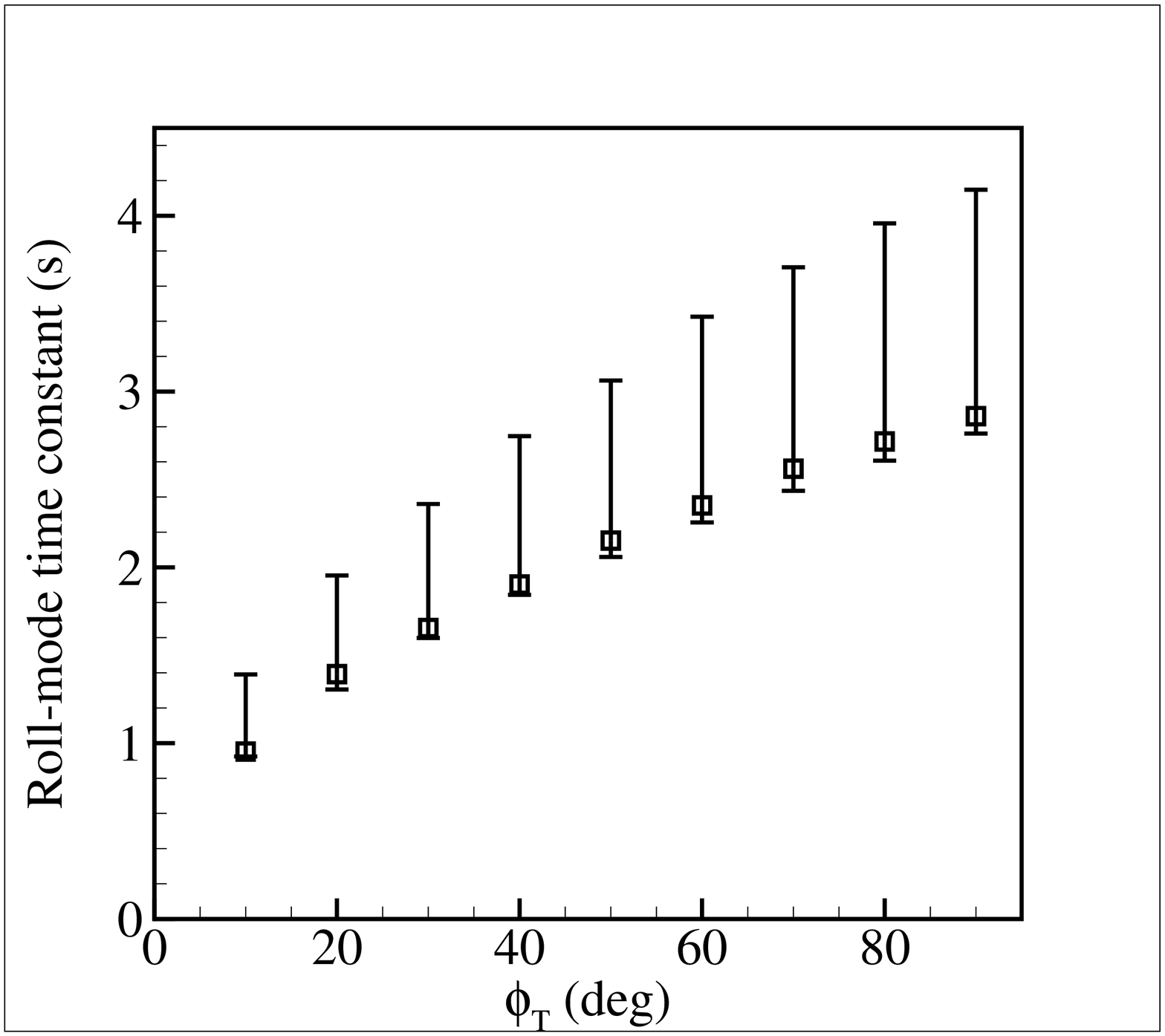}}
      \subfigure[]
       {\label{MC2b}
       \includegraphics[height=55mm]{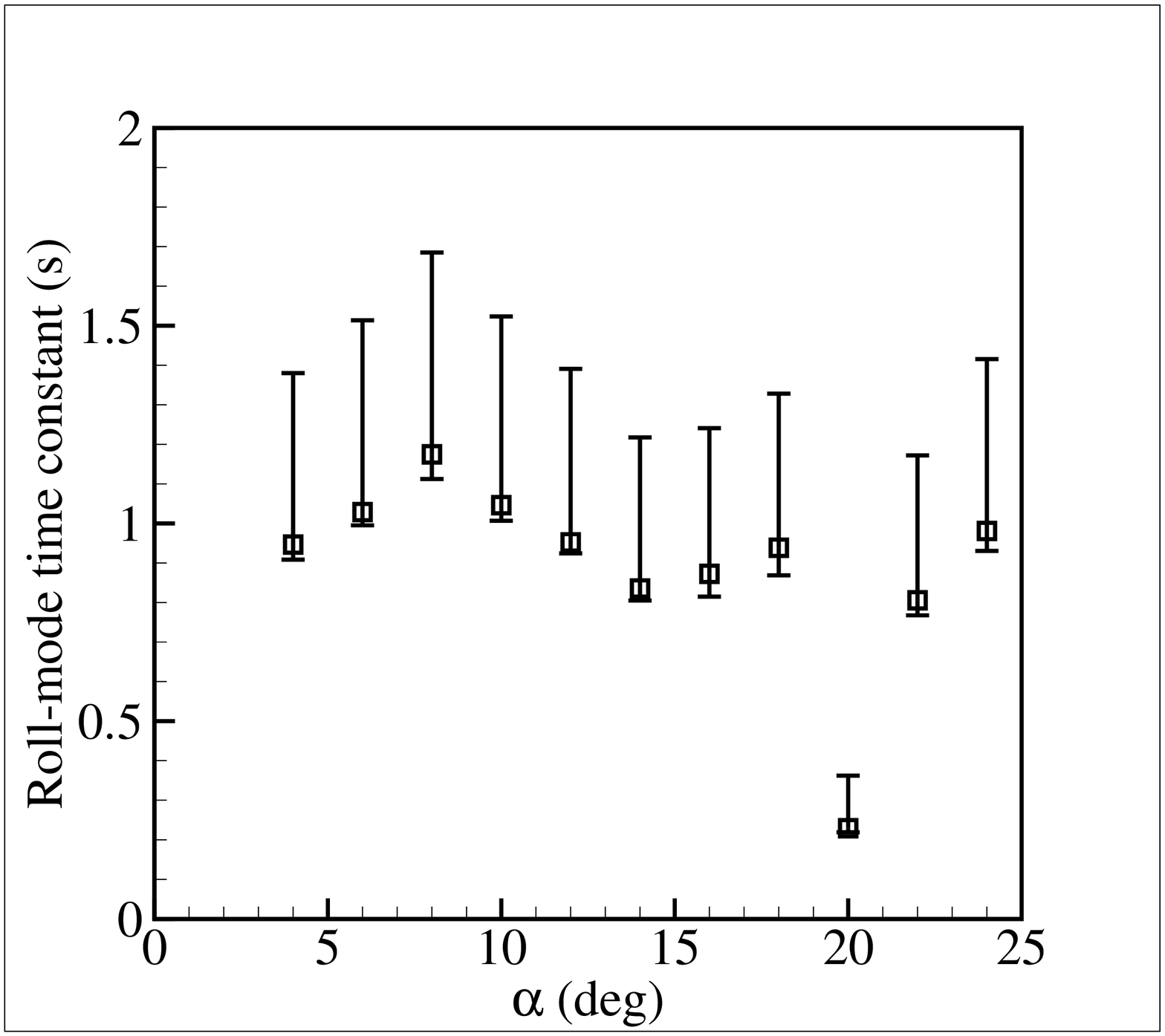}}\\
     \caption{Roll-mode time constant error bar where (a)  $\alpha=12$deg; (b) $\phi=10$deg.}\label{MC2}
     \end{figure}

 A Monte Carlo simulation was conducted to examine the effect of the varying plasma actuations on the roll control. The plasma-induced momentum is approximated with $\pm 20\%$ uncertainty about the nominal values $C_{l_p}$ in Fig. \ref{Clp}. Figures \ref{MC1}-\ref{MC2} show the Monte Carlo simulation results. The roll moment coefficient is randomly chosen between 0.8$C_{l_p}$ and 1.2$C_{l_p}$ in 100 repeated simulations. It can be seen in Fig. \ref{MC1} that overshoots could appear for varied plasma-induced momentum. The phase line slightly changes as well. In addition, the rising time of the control has been affected.

     Figure \ref{MC2} quantitatively shows the changes. The roll-mode time constants (the time needed to reach $63.2\%$ of the target roll angle) for different roll targets (from 10deg to 90deg) at $\alpha=12$deg are shown in Fig. \ref{MC2a}. According to MIL-F-8785C specification of flying quality, the maximum roll-mode time constant is 10s for level 3 flights of light airplanes. Figure \ref{MC2} shows that this flying quality section can be satisfied with the proposed bang-bang controller and plasma actuators. Figure \ref{MC2b} shows the roll-mode time constants for 10 roll degree at various angles of attack between 4deg and 24deg. It can be seen that the roll control is most effective at a high angle of attack around 20deg, where the flow separation has been delayed with plasma actuation. The flying qualities ($<10$s) are still satisfied. 

\begin{figure}
     \centering
     \subfigure[]
       {\label{resultc0}
       \includegraphics[height=40mm]{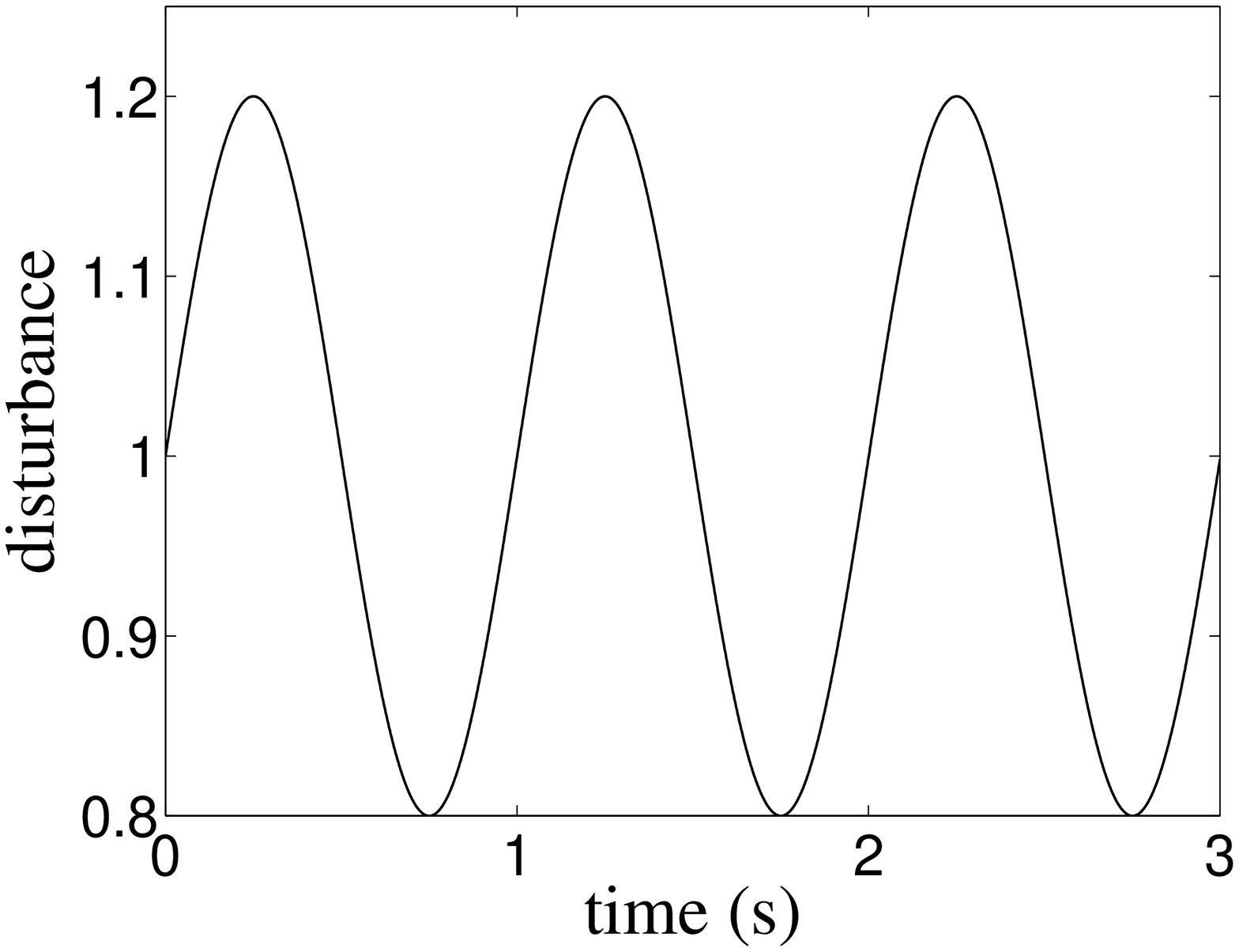}}     
     \subfigure[]
       {\label{resultc1}
       \includegraphics[height=40mm]{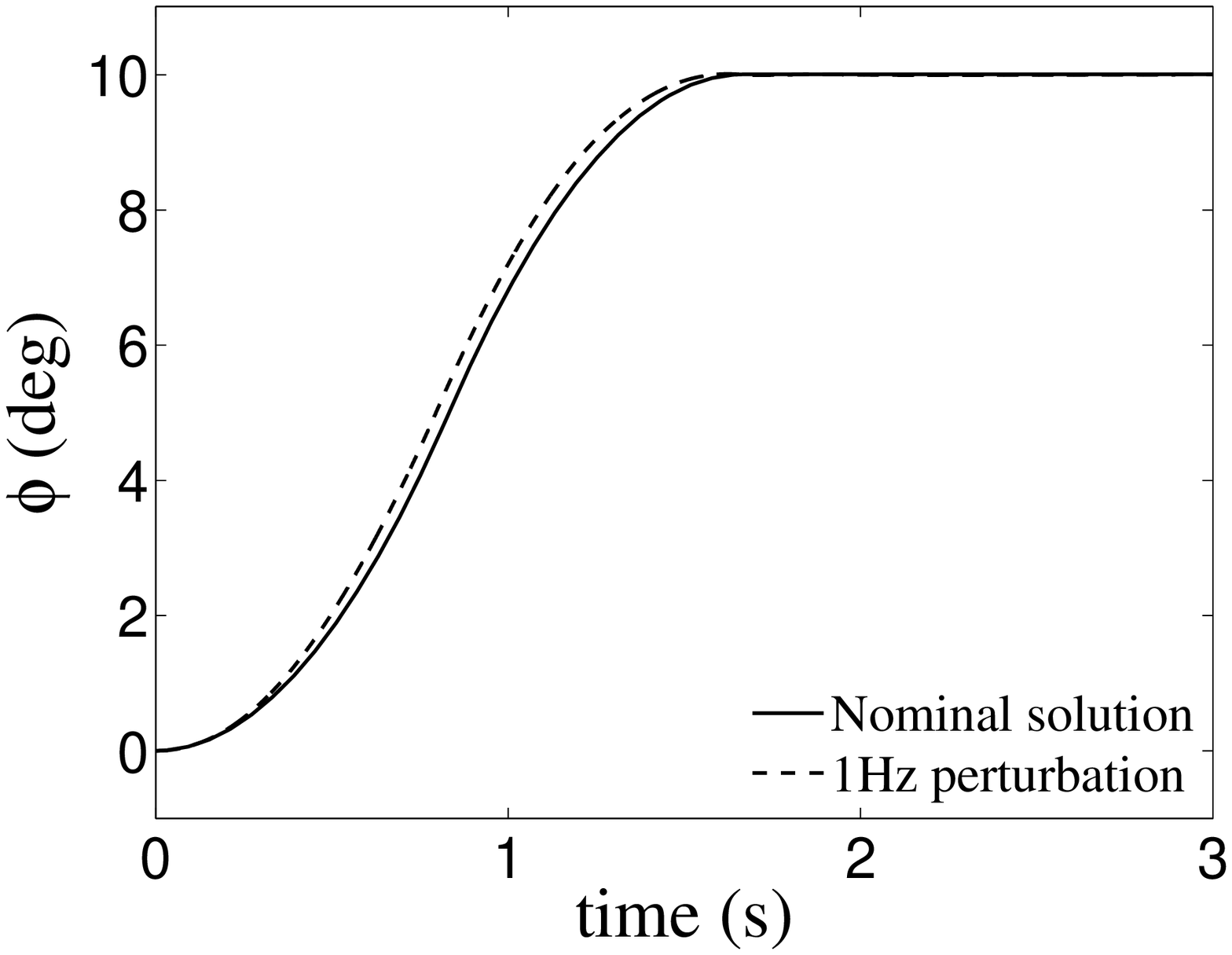}}
     \subfigure[]
       {\label{resultc2}
       \includegraphics[height=40mm]{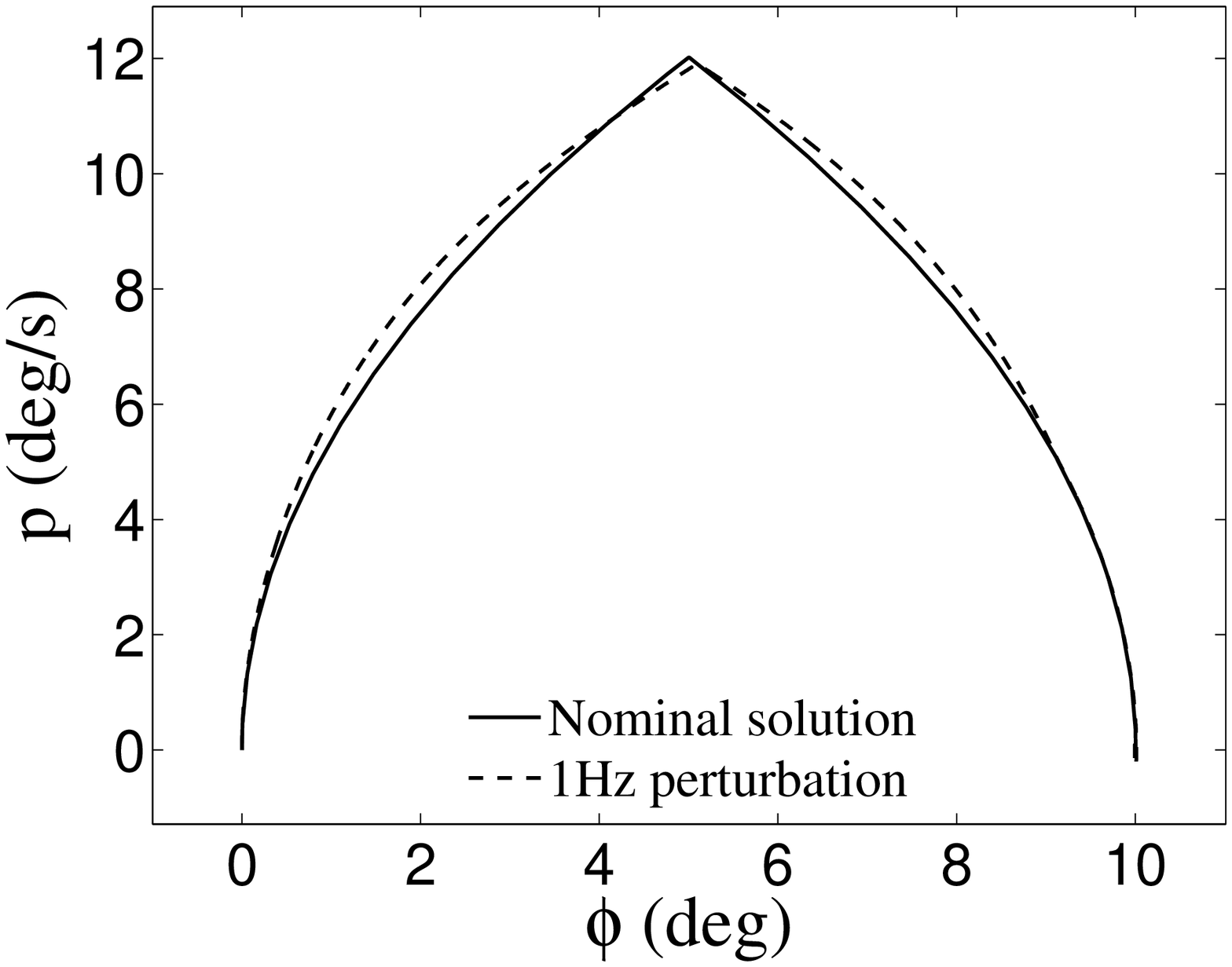}}\\       
       
     \subfigure[]
       {\label{resultd0}
       \includegraphics[height=40mm]{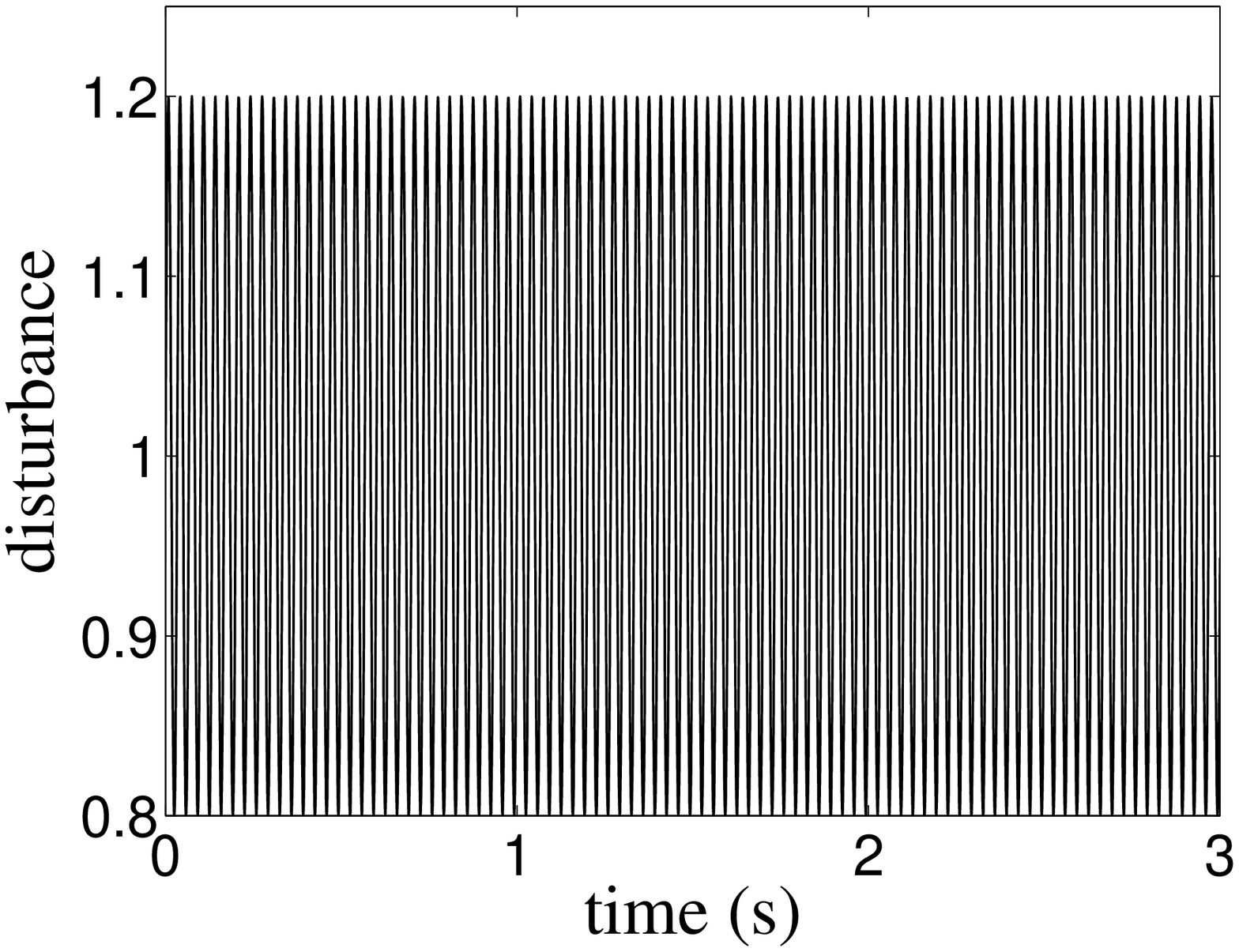}}     
     \subfigure[]
       {\label{resultd1}
       \includegraphics[height=40mm]{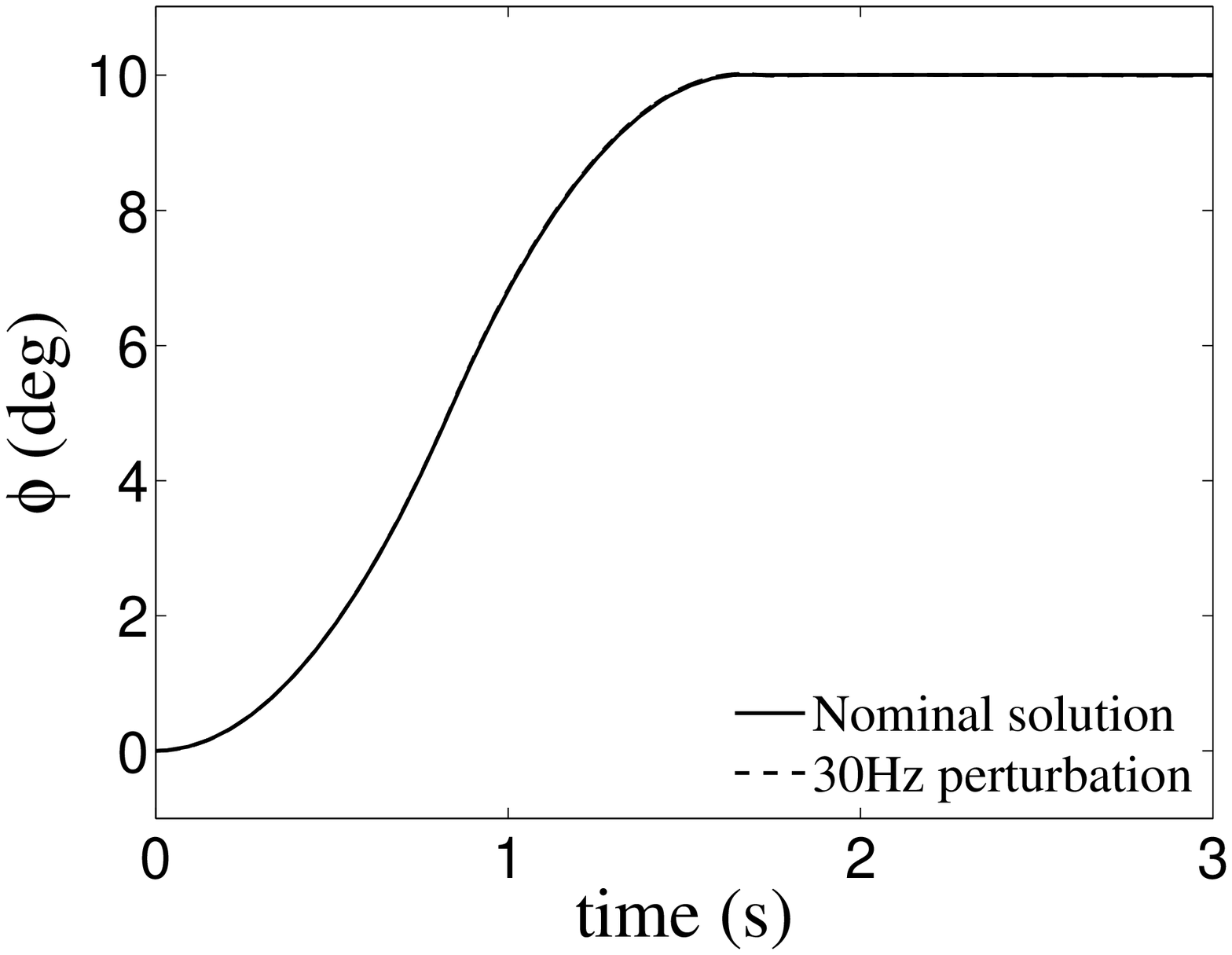}}
     \subfigure[]
       {\label{resultd2}
       \includegraphics[height=40mm]{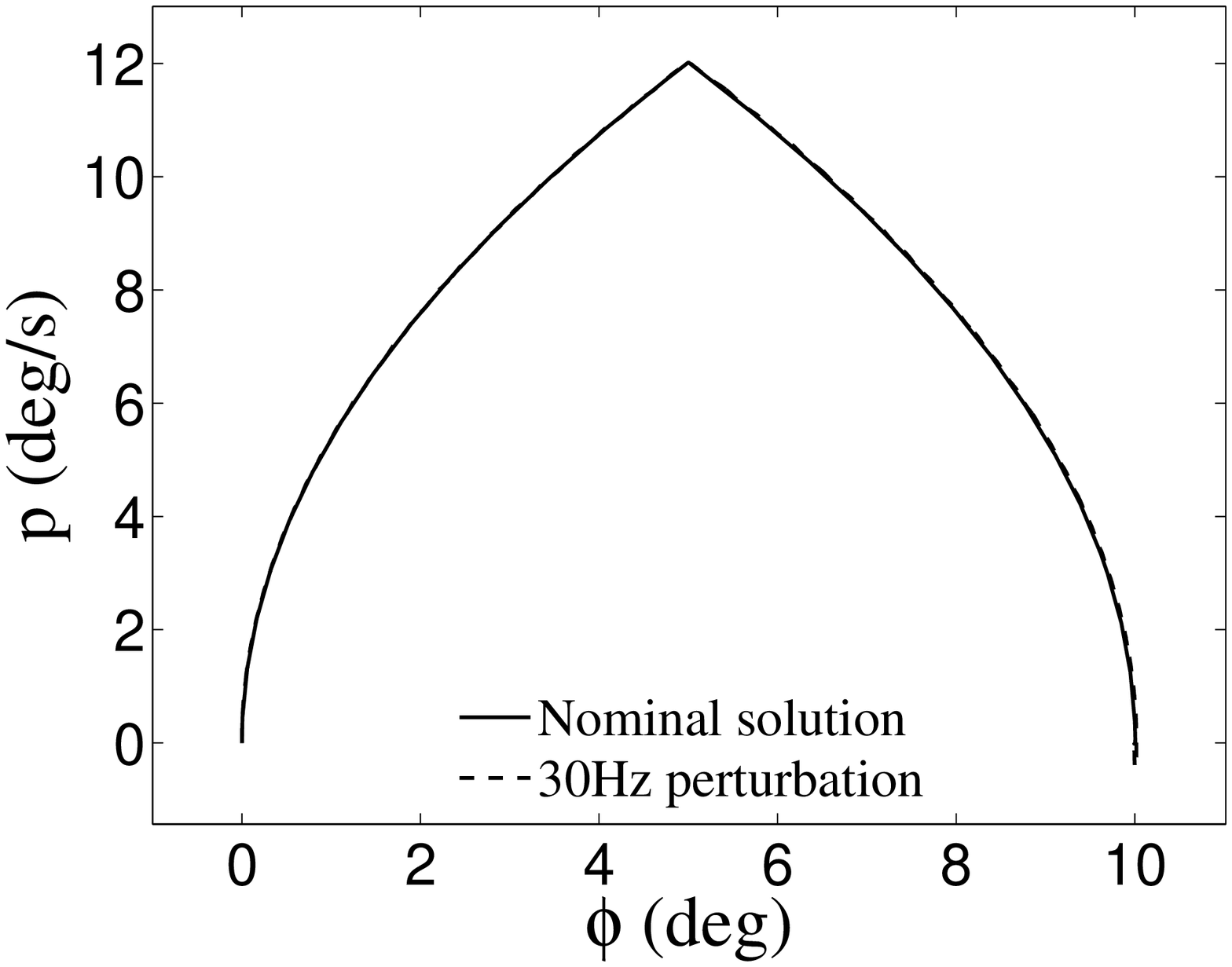}}
     \caption{Simulation results of roll control from 0deg to 10deg at $\alpha=12$deg, where (a)(d) 20$\%$ variance of plasma-induced roll moments at 1Hz and 30Hz, (b)(e) the corresponding roll angle dynamics, and (c)(f) the related phase lines.}
   \end{figure}

It is more practical to regard the almost $20\%$ aerodynamic variance as an instantaneous disturbance to the controlled system. The Fourier spectrum of the results in Fig. \ref{Plasma_velocity} largely lies between 1Hz and 30Hz. As a result, the plasma-induced roll moments are varied at these two frequencies (Fig. \ref{resultc0},\ref{resultd0}) to verify the robustness of the proposed control method. Figures \ref{resultc1}-\ref{resultc2} show that slight change can be found in roll dynamics and phase line ($<5\%$) for the 1Hz perturbation case. Figure \ref{resultd1}-\ref{resultd2} suggests the control method is insensitive to the relatively high frequency disturbance at 30Hz. Almost no change can be found in the roll dynamics and phase line. As a result, the proposed bang-bang control is robust and can address practical issues of plasma actuators. 

     \begin{figure}
     \centering
     \includegraphics[width=100mm]{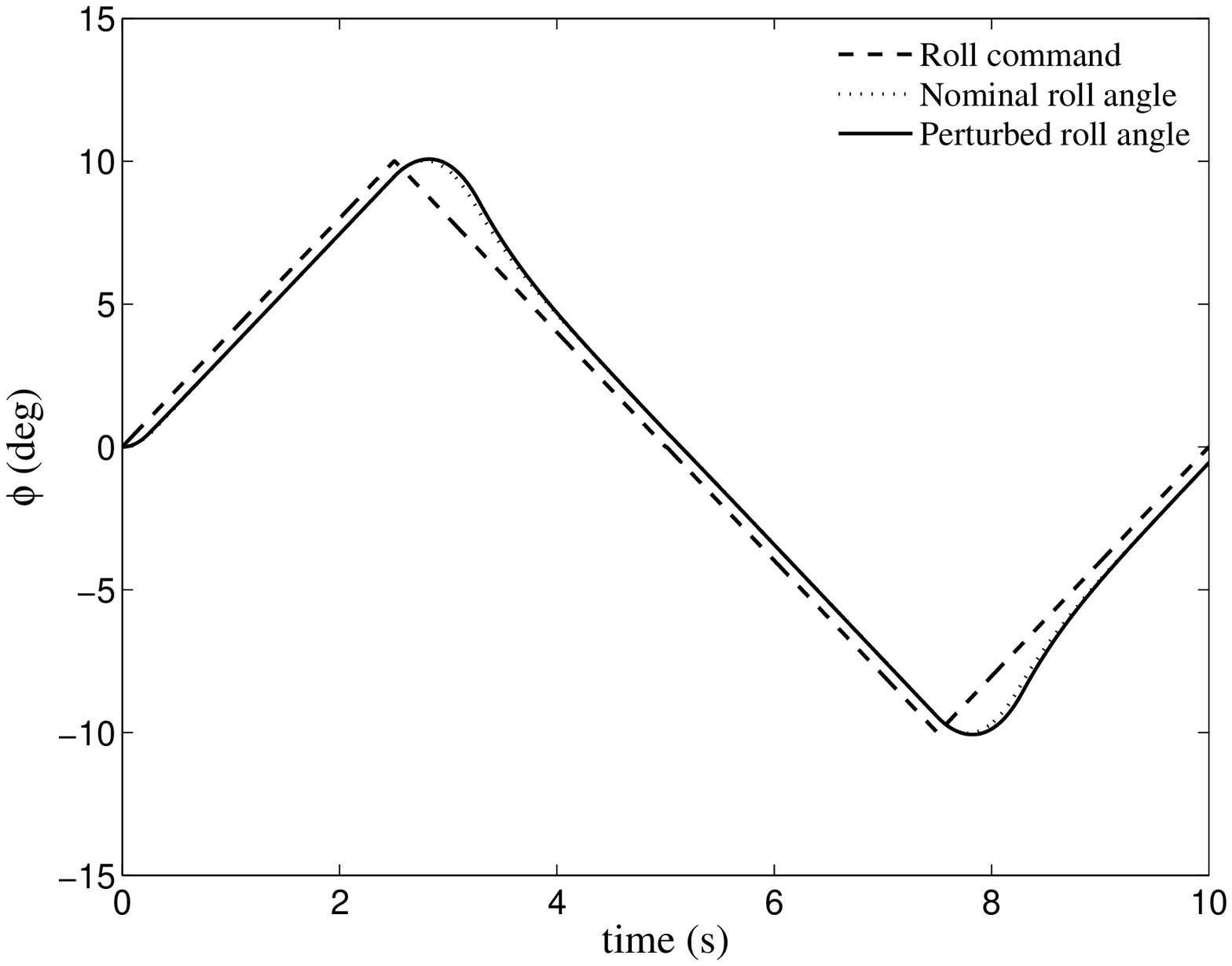} 
        \caption{Roll angle dynamics for a series of roll commands.}\label{TVphi}
     \end{figure}  
At last, Fig. \ref{TVphi} shows the roll dynamics of the airfoil commanded by a series of roll commands. The nominal roll angles with constant plasma-induced roll moments satisfactorily track the roll commands with less than 0.5s time delay. On the other hand, the perturbed roll angles with $20\%$ variances of moments at 1Hz also follow the roll commands well, suggesting the good performance of the proposed control method.

\section{Summary}
The main contribution of this Note is to integrate flight control with active flow control using plasma actuators. The bang-bang control method has been proposed for plasma actuators, taking account of practical issues, such as limited actuation states with instantaneously varied aerodynamic control performance. Flow control effects have been examined in wind tunnel experiments, which show that the plasma authority for flow control is limited. Flow control effects are only obvious at pitch angles near stall. However, flight control simulations suggest that, using the proposed optimal control method, even those small plasma-induced roll moments can satisfactorily fulfill the maneuver tasks and meet flight quality specifications. In addition, the disturbance from volatile plasma-induced roll moments can be rejected well. Hence, the proposed bang-bang control method is a promising candidate of control design methodology for plasma actuators. The ongoing and future works include airfoil flight control in wind tunnel and final flight tests.

\bibliographystyle{aiaa}
\bibliography{DBDroll}

\end{document}